\documentclass[sigconf]{acmart}

\setcopyright{none} 
\acmYear{2017}

\author{TonTon Hsien-De~Huang \\
Leopard Mobile Inc. (Cheetah Mobile Taiwan Agency) \\
Department of Computer Science and Information Engineering, National Cheng Kung University \\
TonTon@TWMAN.ORG
}

\begin{document}
\title{Hunting the Ethereum Smart Contract: Color-inspired Inspection of Potential Attacks}

\begin{abstract}
Blockchain and Cryptocurrencies are gaining unprecedented popularity and understanding. Meanwhile, Ethereum is gaining a significant popularity in the blockchain community, mainly due to the fact that it is designed in a way that enables developers to write smart contract and decentralized applications (Dapps). This new paradigm of applications opens the door to many possibilities and opportunities. However, the security of Ethereum smart contracts has not received much attention; several Ethereum smart contracts malfunctioning have recently been reported. Unlike many previous works that have applied static and dynamic analyses to find bugs in smart contracts, we do not attempt to define and extract any features; instead we focus on reducing the expert's labor costs. We first present a new in-depth analysis of potential attacks methodology and then translate the bytecode of solidity into RGB color code. After that, we transform them to a fixed-sized encoded image. Finally, the encoded image is fed to convolutional neural network (CNN) for automatic feature extraction and learning, detecting compiler bugs of Ethereum smart contract.
\end{abstract}




\maketitle
\section{Introduction}
Since Satoshi Nakamoto published the article "Bitcoin: A Peer-to-Peer Electronic Cash System" in 2008 \cite{bitcoin}, and after the official launch of Bitcoin in 2009, technologies such as blockchain and cryptocurrency have attracted attentions from academia and industry. At present, the technologies have been applied to many fields such as medical science, economics, Internet of Things \cite{surveyvulnerability}. Since the launch of Ethereum (Next Generation Encryption Platform) \cite{BitcoinMagazine} with Vital contract function proposed by Vitalik Buterin in 2015, lots of attention has been obtained on its dedicated cryptocurrency Ether, smart contract, blockchain and its decentralized Ethereum Virtual Machine (EVM). The main reason is that its design method provides developers with the ability to develop Decentralized apps (Dapps), and thus obtain wider applications. A new application paradigm opens the door to many possibilities and opportunities. 

Smart contracts are stored in blocks in order to assist in the negotiation and implementation of the verification contract. The New York Times stated that it distributes and pays the use of public computers that constitute the Internet by many users through the use of Ethereum \cite{NYTIMES}; Bloomberg Businessweek claims that it is a software that is shared to everyone and cannot be tampered \cite{Bloomberg}. Smart contracts can interact with the database at a low cost, resulting in a decentralized autonomous organization \cite{Economist} made up of Ethereum; at the same time, because it is transparent, the smart contract can prove that its claimed function is real, eg. virtual casinos can prove its fairness \cite{Ledger}. However, the security of smart contracts has not received the same attention, and because of its openness, if there are vulnerability in the contract, anyone can immediately see it, but it cannot be stopped immediately. The DAO vulnerabilities that worth more than \$50 million US dollars is an example \cite{Daovulnerability, thedao}. And it raises a similar error through static analysis. There are many vulnerabilities in the Ethereum smart contract, including "Call Stack Attack Vulnerability" \cite{callstack}, "Time Dependance Vulnerability" \cite{Breaking} and various other known vulnerabilities \cite{surveyvulnerability}. These development defects caused thousands of dollars in losses. Compared to traditional programming which can be patched if an error is detected, the smart contract on the blockchain is irreversible and immutable; regardless of its popularity or how much it is worth, the vulnerability cannot be repaired. Hence it is important to further determine the correctness of smart contracts before deployment.

At present, the most widely used security analysis technology is static analysis. However, static analysis needs to be preceded by expert manpower through the analysis of the source code and the acquisition of key code execution sections as a feature to check whether they may result in malicious trading behavior. The static analysis method is less applicable because of the irreversible and immutable nature of the development of smart contracts. Therefore, we have developed a color representation for translating the bytecode of solidity (Ethereum smart contract develop language) into RGB color code and transform them to a fixed-sized encoded image. After that, we also implement a Convolutional Neural Network (CNN) model to perform single-label classification of solidity bytecode images (without extracting features from the solidity source code manually in advance). 
The models we have tried (AlexNet, GoogleNet, and Inception-v3) have achieved good results in our experiments in addition to the good results achieved with the ImageNet Large-Scale Visual Recognition Challenge (ILSVRC); however, the solidity compilation vulnerability usually does not happen alone, so we further optimize it by using multi-label classification of solidity bytecode images through Transfer Learning, which takes only 1.5 seconds per analysis and obtains more accurate multi-label classification results. Because of the inability of smart contracts to be repaired after deployment, we hope that with the tools we have developed, we can provide developers with early detection of relevant vulnerabilities to repair and reduce smart contracts without deploying source code before deployment to reduce the security issues.

\section{Related Work}
A very recent study conducted analysis on nearly one million contracts; among them, 34,200 contracts are vulnerable and nearly 4,000 contracts are practically exploitable. Smart contracts can hardly be adjusted once deployed. To achieve secure contracts, the key step is to have a thorough security examination before deployment \cite{finding}. Loi Luu et al, considered that it is more accurate than the dynamic analysis if the results of the path-by-path method are inferred by static analysis, especially for Ethereum. The uncertainty and complexity of the blockchain behavior makes it difficult to simulate the execution environment. It is the biggest issue for dynamic analysis. Therefore, they developed an execution tool for symbols, called Oyente, to discover potential vulnerability in Ethereum's smart contracts\cite{oyente}.

Each symbolic path has a path condition which is a formula over the symbolic inputs built by accumulating constraints which those inputs must satisfy in order for execution to follow that path. A path is infeasible if its path condition is unsatiable. Otherwise, the path is feasible. Among 19366 existing Ethereum smart contracts, Oyente flags 8833 of them as vulnerable, including the The DAO which led to a \$50 million US dollar loss in June 2016 \cite{ccs2016}. IBM Research presents a framework (ZEUS) to verify the correctness and validate the fairness of smart contracts. ZEUS abstract interpretation and symbolic model checking, along with the power of constrained horn clauses to quickly verify contracts for safety \cite{zeus}. At the same time, they have analyzed almost 22.4K smart contracts，and discover 94.6\% of contracts (containing cryptocurrency worth more than \$0.5 billion) are vulnerable. Microsoft Research, Inria, and Harvard University also have developed a dependent types and monadic effects framework F* (a functional programming language aimed at program verification), and automated queries to statically verify properties on EVM bytecode and Solidity sources \cite{securify}.

In summary, currently, most of the Ethereum security research still rely on labor-intensive examination and focus only on analyzing the control flow graph (CFG) or symbolic execution of the smart contract. so as to determine whether the program under test is causing malicious transaction behavior. This is insufficient in identifying the security flaws in codes. However, Machine Learning/Deep Learning (ML/DL) already has a wide range of applications, especially in security problems, such as spam filtering, botnet detection, and malware classification. Mean in while our proposed system is designed particularly for the security examination of contracts with the minimum labor cost.

\section{Our Proposed Methodology}
We have developed a color representation for translating the bytecode of solidity into RGB color code and transform them to a fixed-sized encoded image. After that, the encoded image is fed to convolutional neural network for automatic feature extraction and learning (without extracting features from the solidity source code manually in advance.), reducing the expert's labor costs. Such translation is also featured by the fact that more complex information in the solidity source code can be preserved in the color image with 16777216 colors (each sampling with 24 bit pixels) compared to the gray scale image with only 256 colors (each sampling with 8 bit pixels). With the fully connected network infrastructure of DNN, though it can deal with its large amount of parameters, however, the local receptive fields and shared weights of CNN make it more suited for more complex structure. It not only decreases the amount of parameters, but also reflects the complexity of smart contract, saving the time for huge computation with current method.

\subsection{The Core Technology of Our Methodology}
\begin{figure}[htbp]
	\centering
	\includegraphics[width=3.4in,height=2.9in]{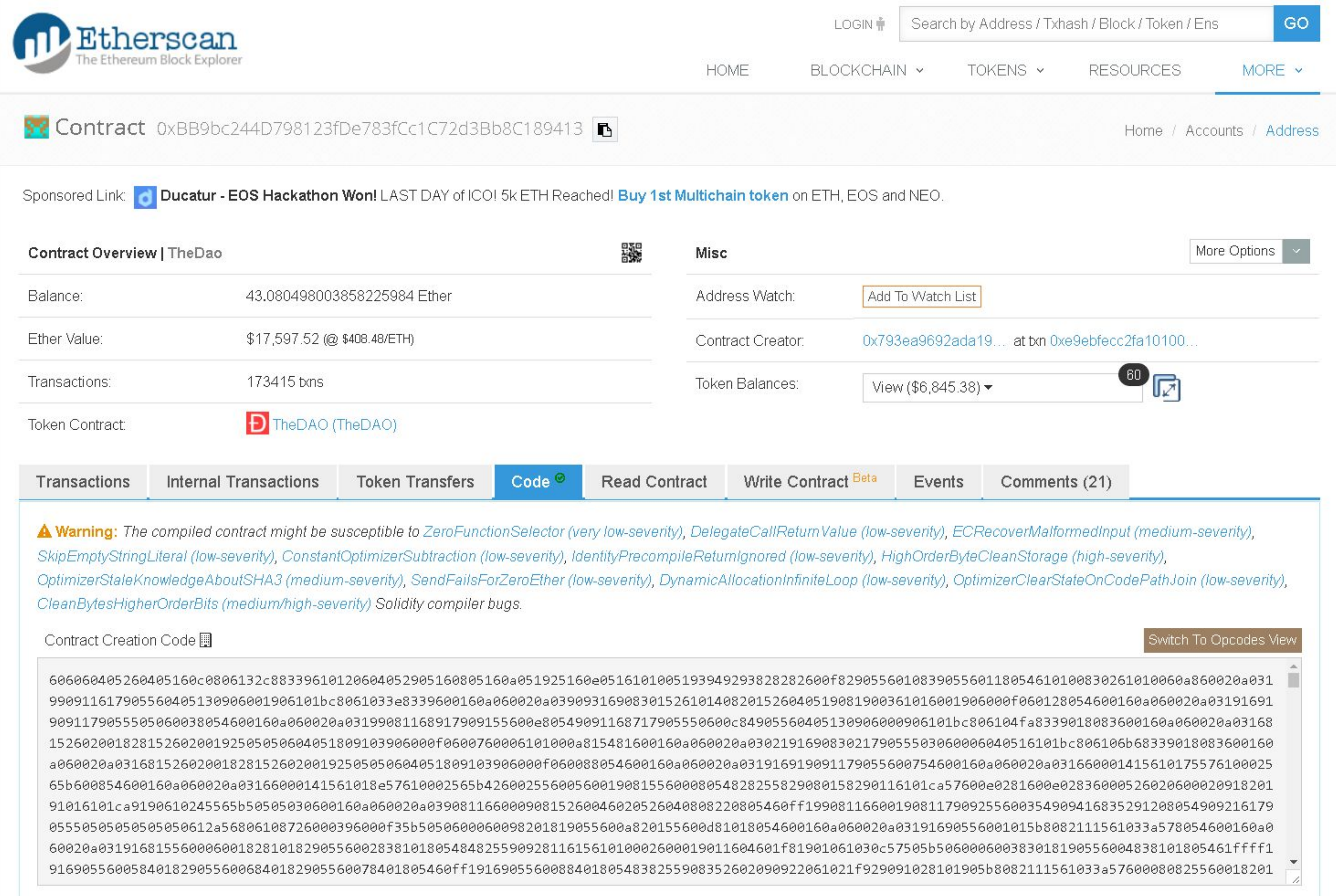}
	\caption{The bytecode of the smart contract example.}\label{fig: F01}
\end{figure}

\begin{figure}[htbp]
	\centering
	\includegraphics[width=3.4in,height=2.5in]{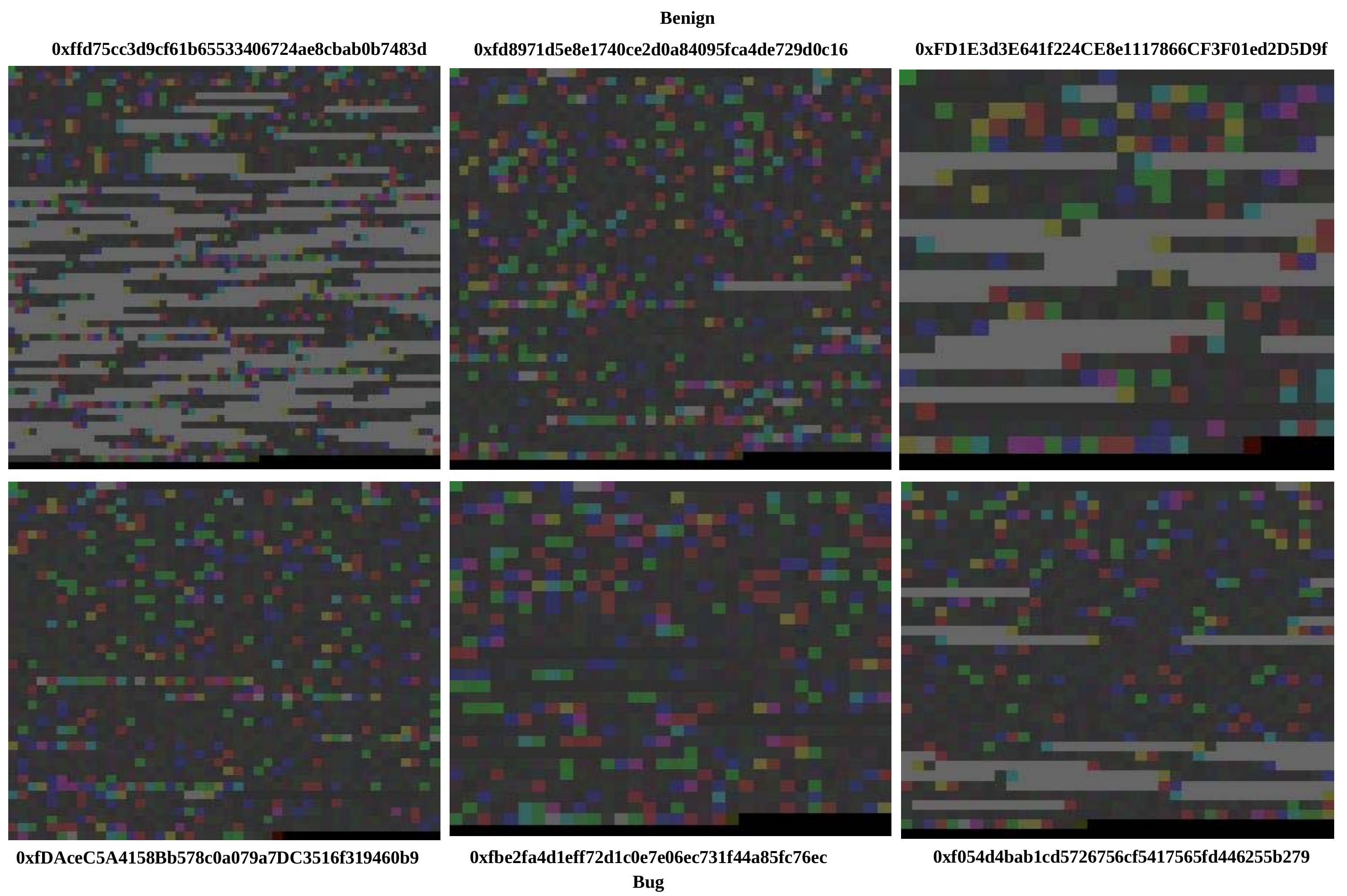}
	\caption{The solidity2jpg of the smart contract example.}\label{fig: F02}
\end{figure}

We now explain further in details. Firstly, taking the smart contract of The DAO (0xBB9bc244D798123fDe783fCc1C72d3Bb8C189413) as an example, its bytecode can be obtained in "https://etherscan.io" (as shown in Fig. \ref{fig: F01}). After that, we perform translating the bytecode of solidity into RGB color code (eg 606060 =(R:96, G:96, B:96), 405260 = (R:64, G:82, B:96), 00357c = ( R: 64, G: 81, B: 96)) and then convert bytecode to rgb color code. After this, we can get color images. Then we input these images to CNN for training the compiler bugs detection model for smart contracts. Fig. \ref{fig: F02} is an example of converting the bytecode of solidity to jpg.

According to our experience, we also found that two approaches might be used to escape our smart contract detection:

\begin{itemize}
\item Since the traditional filter size of CNN is 3*3 or 5*5, the uncorrelated bytecode might become correlated when we transform smart contract into images. The compiler bugs of smart contract may evade the detection by taking advantage of such a mismatch.
\item Meanwhile, smart contract color images are not natural images; instead, they are formed from solidity source code. Thus, the pooling inevitably destroys the contexts and semantics of the malware code, causing the detection inaccuracy.
\end{itemize}

To address the above two issues, we did many experiments with CNN models (includes AlexNet, GoogleNet, and Inception-v3). We found the characteristics of 1x1 convolution in "Network in Network " \cite{NIN}. 1x1 convolution is equivalent to cross-channel parametric pooling layer, and this cascaded cross channel parametric pooling structure allows complex and learnable interactions of cross channel information.

\subsection{The Architecture of Our Methodology}
\begin{figure}[htbp]
	\centering
	\includegraphics[width=3.4in,height=2.7in]{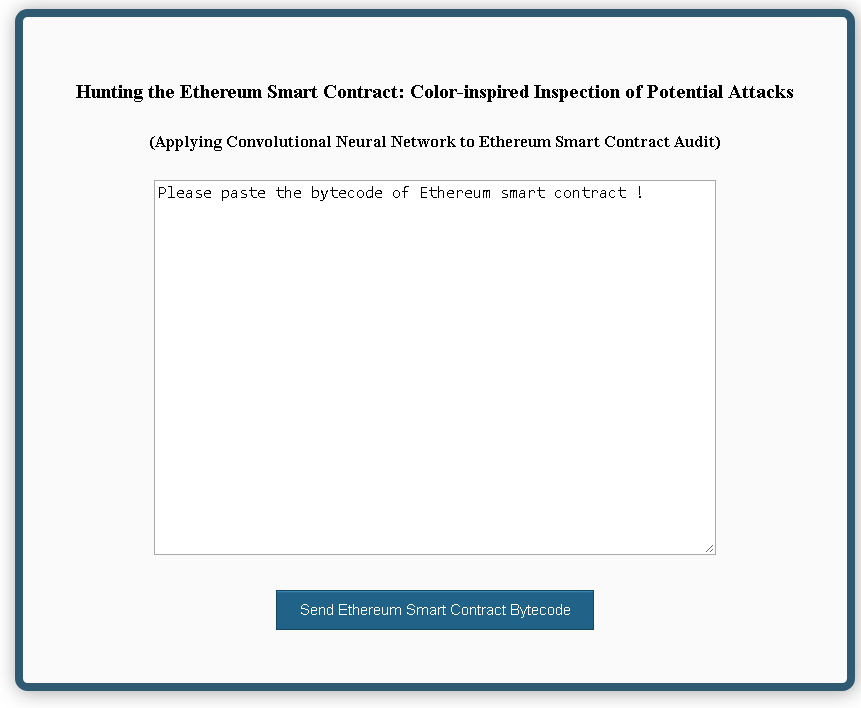}
	\caption{The screenshot of our proof of concept UI.}\label{fig: F03}
\end{figure}

\begin{figure}[htbp]
	\centering
	\includegraphics[width=3.4in,height=2.6in]{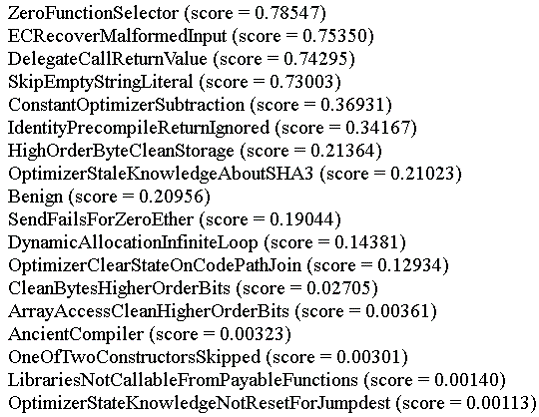}
	\caption{The screenshot of our RESTful API result .}\label{fig: F04}
\end{figure}

Fig. \ref{fig: F03} is a screenshot of our internal proof-of-concept UI, and Fig. \ref{fig: F04} is the result of the passback of our RESTful API (multi-label). A brief description of our system flow chart is shown as follows (steps 1-4 are off-line phases for our internal development and steps 5-6 are online phases, where the user can interact with the system. More specifically, after uploading the bytecode and invokes our RESTful API, the user can obtain the analysis result.):

\begin{itemize}
\item Step 1. Crawling the bytecode of smart contract from etherscan by pyspider (including benign and malicious ones) as sample;
\item Step 2. Transforming the bytecode of smart contract into RGB color code and transforming them to a fixed-sized encoded image;
\item Step 3. After that, the encoded image is fed to convolutional neural network for automatic feature extraction and learning (without extracting features from the solidity source code manually in advance.)
\item Step 4. Finally, once the model has been trained and validated, we deploy it on the backend server.
\item Step 5. Only provide the bytecode of smart contract, we will transform it into smart contract color image.
\item Step 6. After the bytecode of smart contract are all identified, the scanned results will be provided to the users through our visualization tool and public RESTful API.
\end{itemize}

However, we also found more key problems. Because the gap between the number of normal contract and the number of vulnerable contract samples is very large, and currently there is only 17 Solidity known bugs such as "optimizerStateKnowledgeNotResetForJumpdest", "ArrayAccessCleanHigherOrderBits", and " AncientCompiler" etc., and their gap is also very large. We need to collect those imbalanced data and we need sufficient training data in this domain. It means that a picture will be defined in multiple categories. Therefore, based on our previous research \cite{TonTon}, we re-implement Inception-v3 to multi-label classification through Transfer Learning. For example, 0xbBCf10D6bc180172d8d352BE5bBCfB814E8f3474 is at the same time with Solidity bugs including SolidFunSelectSelector, DelegateCallReturnValue, ECRecoverMalformedInput and SkipEmptyStringLiteral.

\begin{figure}[htbp]
	\centering
	\includegraphics[width=3.2in,height=2.5in]{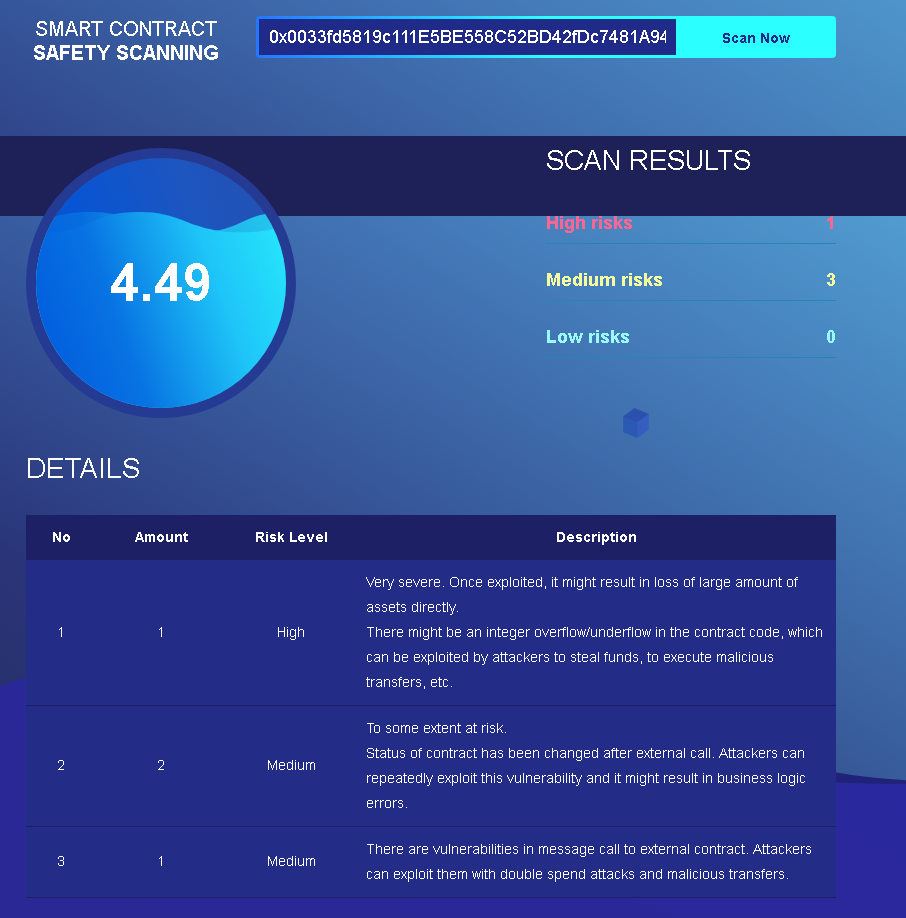}
	\caption{The screenshot of our proof of concept scan result.}\label{fig: F05}
\end{figure}

\section{Experiment Result}

We run it on a 64-bit Ubuntu 14.04, and hardware setting are 128 GB DDR4 2400 RAM and Intel(R) Xeon(R) E5-2620 v4 CPU, NVIDIA TITAN V, TITAN XP and GTX 1080 GPUs; more specifically, the software setting is the nvidia-docker tensorflow:18.04-py2 on NVIDIA cloud. The research results and the data will be found on our website http://R2D2.TWMAN.ORG.

We first crawled from the etherscan.io for the verified contract information from Jan. 2018 to Apr. 2018. The smart contract message on this website represents that the ethereum mainnet has been verified. We further translate them into color images and feed them to CNN for training.
\begin{itemize}
\item AlexNet
\begin{itemize}
\item 100 epoch, LR: 0.01, accuracy: 83.85\%, loss: 0.34
\item 250 epoch, LR: 0.01, accuracy: 86\%, loss: 0.39
\end{itemize}
\item GoogleNet
\begin{itemize}
\item 100 epoch, LR: 0.01, accuracy: 86\%, loss: 0.32
\item 250 epoch, LR: 0.01, accuracy: 90\%, loss: 0.53
\end{itemize}
\item Inception-v3
\begin{itemize}
\item 100 epoch, LR: 0.001, accuracy: 97.10\%, loss: 0.1345
\item 100 epoch, LR: 0.0001, accuracy: 96.52\%, loss: 0.0935
\item 500 epoch, LR: 0.001, accuracy: 97.39\%, loss: 0.1876
\item 500 epoch, LR: 0.0001, accuracy: 83.76\%, loss: 1.08445
\end{itemize}
\end{itemize}

After that, we collected from May. 2018 to Jun 2018 from etherscan.io as our test dataset with Inception-v3. We calculate metrics such as accuracy, precision and recall and show the results as follows.

\begin{itemize}
\item 100 epoch, LR 0.001, accuracy: 95.44\%, precision: 95.42\%, recall: 98.5\%
\item 100 epoch, LR 0.0001, accuracy: 95.61\%, precision: 95.26\%, recall: 98.9\%
\item 500 epoch, LR 0.001, accuracy: 95.85\%, precision: 95.44\%, recall: 99.04\%
\item 500 epoch, LR 0.0001, accuracy: 83.65\%, precision: 86.48\%, recall: 91.46\%
\end{itemize}

On the other hand, since the solidity compile bugs of smart contracts have the characteristics of multi-label, we initially identify if there are compile bugs. If the analysis results are malicious, we use transfer learning for multi-label. After training and inference, the results obtained are shown in Figure 4. Fig. \ref{fig: F05} is a screenshot of our public proof-of-concept scan result.

\section{Conclusion}

According to our data, there are 1800 new smart contract produced on ethereum main net per day. Amongst less than 30\% is verified by etherscan. Meanwhile according to the third party blockchain evaluation system, Rating Token (https://ratingtoken.io) and coinschedule, in 2018, there has been 11.75 billion US dollars financed by ICO projects. One of the key factors for the success of a smart contract for each project is the existence of a loophole; Our goal is to optimize the amount of parameters, network structure, and release automated verification tools and public RESTful API. The above experiment results demonstrate that our proposed system can have accurate security analysis on contracts with very limited labor cost.

\bibliographystyle{ACM-Reference-Format}

\begin{thebibliography}{99}
\bibitem{bitcoin}
S. Nakamoto. Bitcoin: A peer-to-peer electronic cash system. http://bitcoin.org/bitcoin.pdf.

\bibitem{surveyvulnerability}
X. Li, P. Jiang, T. Chen, X. Luo, and Q. Wen, "A survey on the security of blockchain systems," Future Generation Computer Systems,arxiv: 1802.06993

\bibitem{BitcoinMagazine}
Vitalik Buterin, “A Next-Generation Cryptocurrency and Decentralized Application Platform,” Bitcoin Magazine, [Online]. Available: https://bitcoinmagazine.com/articles/ethereum-next-generation-cryptocurrency-decentralized-application-platform-1390528211/

\bibitem{NYTIMES}
“Ethereum, a Virtual Currency, Enables Transactions That Rival Bitcoin’s,” Nathaniel Popper for the New York Times, [Online]. Available: https://www.nytimes.com/2016/03/28/business/dealbook/ethereum-a-virtual-currency-enables-transactions-that-rival-bitcoins.html

\bibitem{Bloomberg} 
“This Is Your Company on Blockchain,” Bloomberg Businessweek, [Online]. Available: https://www.bloomberg.com/news/articles/2016-08-25/this-is-your-company-on-blockchain

\bibitem{Economist}
“The great chain of being sure about things,” The Economist, [Online]. Available: https://www.economist.com/briefing/2015/10/31/the-great-chain-of-being-sure-about-things

\bibitem{Ledger} 
P. J. Piasecki, "Gaming Self-Contained Provably Fair Smart Contract Casinos," 2016, vol. 1, p. 12, 2016.

\bibitem{Daovulnerability} 
V. Buterin. Critical update re: Dao vulnerability. https://blog.ethereum.org/2016/06/17/critical- update-re-dao-vulnerability, 2016.

\bibitem{thedao}
"The DAO smart contract," [Online]. Available: http://etherscan:io/address/0xbb9bc244d798123fde783fcc1c72d3bb8c189413\#code

\bibitem{callstack}
"Gas Economics: Call Stack Depth Limit Errors," [Online]. Available: https://github.com/LeastAuthority/ethereumanalyses/blob/master/GasEcon.md\#callstack-depth-limit-errors

\bibitem{Breaking}
"Breaking the House," [Online]. Available: http://martin.swende.se/blog/Breaking the house.html

\bibitem{finding}
I. Nikolic, A. Kolluri, I. Sergey, P. Saxena, and A. Hobor, "Finding The Greedy, Prodigal, and Suicidal Contracts at Scale," arXiv:1802.06038, 2018.

\bibitem{ccs2016}
L. Luu, D. Chu, H. Olickel, P. Saxena, and A. Hobor, "Making smart contracts smarter," in CCS. ACM, 2016, pp. 254¡V269.

\bibitem{oyente}
"Oyente: An Analysis Tool for Smart Contracts," [Online]. Available: https://github.com/melonproject/oyente

\bibitem{zeus}
S. Kalra, S. Goel, M. Dhawan, and S. Sharma, "Zeus: Analyzing safety of smart contracts," in 25th ISOC Symposium on Network and Distributed System Security (NDSS'18), 2018.

\bibitem{securify}
K. Bhargavan, A. Delignat-Lavaud, C. Fournet, A. Gollamudi, G. Gonthier, N. Kobeissi, et al., "Formal Verification of Smart Contracts: Short Paper," presented at the Proceedings of the 2016 ACM Workshop on Programming Languages and Analysis for Security, Vienna, Austria, 2016.



\bibitem{NIN}
M. Lin, C. Qiang, and Y. Shuicheng, "Network In Network," 2th International Conference on Learning Representations (ICLR), Banff, Canada, 2014.

\bibitem{TonTon}
TonTon H.-D. Huang, Chia-Mu Yu, and Hung-Yu Kao, "R2-D2: Color-Inspired Convolutional Neural Network (CNN)-based Android Malware Detection", arXiv:1705.04448.
\end{thebibliography}

\end{document}